\documentclass[prb,superscriptaddress,amsfonts,twocolumn,showpacs,floatfix]{revtex4}
\usepackage{amsmath}
\usepackage{amsfonts}
\usepackage{amssymb}

\usepackage{txfonts}

\usepackage{bm}
\usepackage{tabularx}
\usepackage{graphicx,color}
\usepackage{textcomp}

\def\Vec#1{\bm{#1}}
\def\Hc2{H_\mathrm{c2}}
\def\Tc{T_\mathrm{c}}
\def\Ic{I_\mathrm{c}}
\def\214{\mathrm{Sr_2RuO_4}}

\newcommand{\GAmma}{\symbol{'000}}
\newcommand{\PHi}{\symbol{'010}}

\begin{document}

\title{Angular dependence of the upper critical field of Sr$_2$RuO$_4$
}
\author{S. Kittaka}
\affiliation{Department of Physics, Graduate School of Science, Kyoto University, Kyoto 606-8502, Japan}
\author{T. Nakamura}
\affiliation{Department of Physics, Graduate School of Science, Kyoto University, Kyoto 606-8502, Japan}
\author{Y. Aono}
\affiliation{Department of Physics, Graduate School of Science, Kyoto University, Kyoto 606-8502, Japan}
\author{S. Yonezawa}
\affiliation{Department of Physics, Graduate School of Science, Kyoto University, Kyoto 606-8502, Japan}
\author{K. Ishida}
\affiliation{Department of Physics, Graduate School of Science, Kyoto University, Kyoto 606-8502, Japan}
\author{Y. Maeno}
\affiliation{Department of Physics, Graduate School of Science, Kyoto University, Kyoto 606-8502, Japan}

\date{\today}
\begin{abstract}
One of the remaining issues concerning the spin-triplet superconductivity of $\214$ is the strong limit of the in-plane upper critical field $\Hc2$ at low temperatures.
In this study, we clarified the dependence of $\Hc2$ on the angle $\theta$ between the magnetic field and the $ab$ plane at various temperatures,
by precisely and accurately controlling the magnetic field direction.
We revealed that, 
although the temperature dependence of $\Hc2$ for $|\theta| \ge 5^\circ$ is well explained by the orbital pair-breaking effect,
$\Hc2(T)$ for $|\theta| < 5^\circ$ is clearly limited at low temperatures.
We also revealed that 
the $\Hc2$ limit for $|\theta| < 5^\circ$ is present not only at low temperatures, but also at temperatures close to $\Tc$.
These features may provide additional hints for clarifying the origin of the $\Hc2$ limit.
Interestingly, if the anisotropic ratio in $\214$ is assumed to depend on temperature, 
the observed angular dependence of $\Hc2$ is reproduced better at lower temperature 
with an effective-mass model for an anisotropic three-dimensional superconductor.
We discuss the observed behavior of $\Hc2$ based on existing theories.
\end{abstract}
\pacs{74.25.Dw, 74.70.Pq}

\maketitle
\section{Introduction}

The layered perovskite superconductor $\214$ with the transition temperature $\Tc$ of 1.5~K 
has been extensively studied due to its unconventional pairing state. \cite{Maeno1994Nature,Mackenzie2003RMP}
Knight shift measurements with NMR \cite{Ishida1998Nature,Ishida2001PRB} and with spin-polarized neutron-scattering \cite{Duffy2000PRL} have revealed 
the invariant spin susceptibility across $\Tc$ for $H \parallel ab$, 
which firmly indicates that the spin part of the Cooper-pair state is triplet.
The orbital part is favorably interpreted as odd parity 
based on the measurements of the critical current $\Ic$ through Pb/$\214$/Pb proximity junctions \cite{Jin1999PRB,Honerkamp1998PTP}
and other experiments.\cite{Nelsoni2004Science,Kidwingira2006Science}
These results establish that $\214$ is an odd-parity spin-triplet superconductor.
In addition, the $\muup$SR\cite{Luke1998Nature} and Kerr effect \cite{Xia2006PRL} measurements 
indicate broken time-reversal symmetry in the superconducting state.
The zero-field ground state consistent with all these results is expressed by the vector order parameter, the $d$-vector, 
$\Vec{d}=\Delta_0 \hat{z} (k_x \pm ik_y)$.
However, recent Ru-NMR measurements under very low fields down to 20~mT revealed 
the invariant Knight shift for $H \parallel c$. \cite{Murakawa2004PRL,Murakawa2007JPSJ}
This means $\Vec{d} \parallel ab$~($\perp H$) with the following two possibilities.\cite{Murakawa2004PRL,Murakawa2007JPSJ,Kaur2005PRB}
The $d$-vector can rotate freely in the $ab$ plane,\cite{Annett2008PRB} or
the $d$-vector pointing along the $c$ axis in zero field can flip perpendicular
to the $c$ axis by a small magnetic field along the $c$ axis.\cite{Yoshioka2009JPSJ}
In either case,
the spin of the Cooper pair can be polarized to any field directions at least above 20 mT.

Another unsolved issue in $\214$ is the origin of the strong limit of the upper critical field $\Hc2$,
which occurs when a magnetic field is applied parallel to the $ab$ plane.\cite{Deguchi2002JPSJ}
Similar $\Hc2$ limit is observed in another spin-triplet superconductor UPt$_3$ for $H \parallel c$, \cite{Dijk1993JLTP} 
as shown in the inset of Fig.~\ref{HT}(b).
These $\Hc2$ limits are reminiscent of the Pauli effect, which results from the Zeeman energy of quasiparticles.
However, in spin-triplet superconductors, the Pauli effect contributes to pair-breaking only when $\Vec{d} \parallel H$,
because the spin of the triplet Cooper pairs can be polarized along the field direction when $\Vec{d} \perp H$.
As mentioned above, the $d$-vector of $\214$ is likely to be perpendicular to the magnetic field possibly except at low fields.
This suggests that the Pauli effect should not affect $\Hc2$. 
We note that the $d$-vector of UPt$_3$ for $H \parallel c$ was 
revealed to be perpendicular to the magnetic field ($\Vec{d} \parallel a$) in phase C. \cite{Tou1998PRL}
Therefore, the $\Hc2$ limit observed in UPt$_3$ cannot be attributed to the Pauli effect either. 
The origins of these $\Hc2$ limits have not been clarified yet.

In this paper, we report the dependence of $\Hc2$ of $\214$, determined from the ac susceptibility, 
on the magnetic field direction between the $ab$ plane and $c$ axis.
Because of the large anisotropy of $\Hc2$ in $\214$, 
a small misalignment would lead to a large difference in the value of $\Hc2$,
especially when the field direction is nearly parallel to the $ab$ plane.
Therefore, in this study, we controlled the applied field direction more accurately and precisely than in the previous reports. \cite{Mao2000PRL, Yaguchi2002PRB}
We evaluated the $\Hc2(T)$ curves for different field directions and 
revealed that the $\Hc2$ limit is clearly observed only when 
the angle $\theta$ between the magnetic field and the $ab$ plane is less than 5 degrees.
This $\Hc2$ limit was revealed to occur not only at low temperatures, but also at temperatures close to $\Tc$.
We also identified the angle $\theta$ dependence of $\Hc2$ at several fixed temperatures. 
We found that $\Hc2(\theta)$ is fitted better at lower temperature with an effective-mass model for an anisotropic three-dimensional superconductor.
In addition, we investigated the difference between $\Hc2(\theta)$ for fields in the (100) plane and for fields in the (110) plane.
The difference appears only at low temperatures below roughly 1~K for small $\theta$.
These results allow us to reexamine the origin of the $\Hc2$ limit based on existing theories.

\section{Experimental}

We used single crystals of $\214$ grown by a floating zone method.\cite{Mao2000MRB}
In this paper, we focus on the result obtained from a single crystal with dimensions of approximately 1.0 $\times$ 0.5 mm$^2$ in the $ab$ plane and 0.08 mm along the $c$ axis.
The directions of the tetragonal crystallographic axes of the sample were determined from x-ray Laue pictures.
We shaped the sample so that the side surface of the sample was 10 degrees away from the (100) plane in order to avoid possible anisotropy effects due to surface superconductivity.\cite{Keller1996PRB}
The crystal was annealed in oxygen at 1~atm and 1050~\textdegree{}C for a week to reduce the amount of oxygen deficiencies and lattice defects.
A sharp superconducting transition was observed in the ac susceptibility measurements with the midpoint at $\Tc$=1.503~K.

\begin{figure}
\includegraphics[width=3.2in]{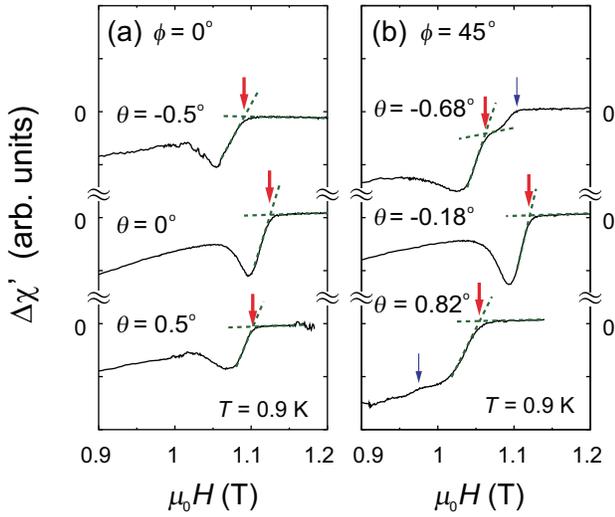}
\caption{
(Color online) 
Field dependence of $\Delta \chi^\prime=\chi^\prime(0.9\ \rm{K})-\chi^\prime(2\ \rm{K})$ at several $\theta$ for (a) $\phi=0^\circ$ and (b) $\phi=45^\circ$.
Here, $\phi$ denotes the azimuthal angle within the $ab$ plane between the magnetic field and the [100] axis and
$\theta$ denotes the angle between the magnetic field and the $ab$ plane.
$\Hc2$ is defined as the intersection of the linear extrapolations in $\Delta\chi^\prime$.
Thick and thin arrows represent $\Hc2$ and the anomaly due to the mosaic structure, respectively.
The dip in $\Delta \chi^\prime$ near $\Hc2$ is attributable to the ordinary peak effect. \cite{Yaguchi2002PRB}
}
\label{raw}
\end{figure}

We measured the ac magnetic susceptibility $\chi_\mathrm{ac}=\chi^\prime-i \chi^{\prime \prime}$ by a mutual-inductance technique using a lock-in amplifier with a frequency of 887 Hz. 
The sample was cooled down to 70~mK with a $^3$He-$^4$He dilution refrigerator.
The ac magnetic field of 20~$\muup$T-rms was applied nearly parallel to the $c$ axis with a small coil.
The dc magnetic field was applied using the ``Vector Magnet'' system, \cite{Deguchi2004RSI}
with which we can control the field direction three dimensionally and precisely. 
The accuracy and precision of the field alignment with respect to the $ab$ plane are better than 0.1 degree and 0.01 degree, respectively.
Owing to the high sensitivity of the pick-up-coil, parasitic background contributes to the $\chi_\mathrm{ac}$ signal.\cite{Kittaka2009JPCS-2}
In order to obtain $\chi_\mathrm{ac}$ contribution only from a superconductivity $\Delta \chi^\prime$, 
we adopt $\Delta\chi^\prime(T,H) = \chi^\prime(T,H)-\chi^\prime(2~\mathrm{K},H)$.
The small deviation of the normal state values from zero indicates a good reliability of the background subtraction. 
We define $\Hc2$ as the intersection between the linear extrapolations of $\Delta \chi^\prime(H)$ in the superconducting and normal states,
as illustrated in Fig.~\ref{raw} with dashed lines. 
The directions of the crystalline axes [100] and [001] with respect to the field direction were calibrated by making use of the anisotropy in $\Hc2$.\cite{Yaguchi2002PRB,Mao2000PRL}
Our highly accurate and precise measurements revealed
that the present sample has a mosaic structure dominated by two domains sharing the [100] axis; 
the [001] axis of one domain is tilted nearly toward the [010] axis by 0.5 degree from the [001] axis of the other part.

\section{Results}

\begin{figure}
\includegraphics[width=3.2in]{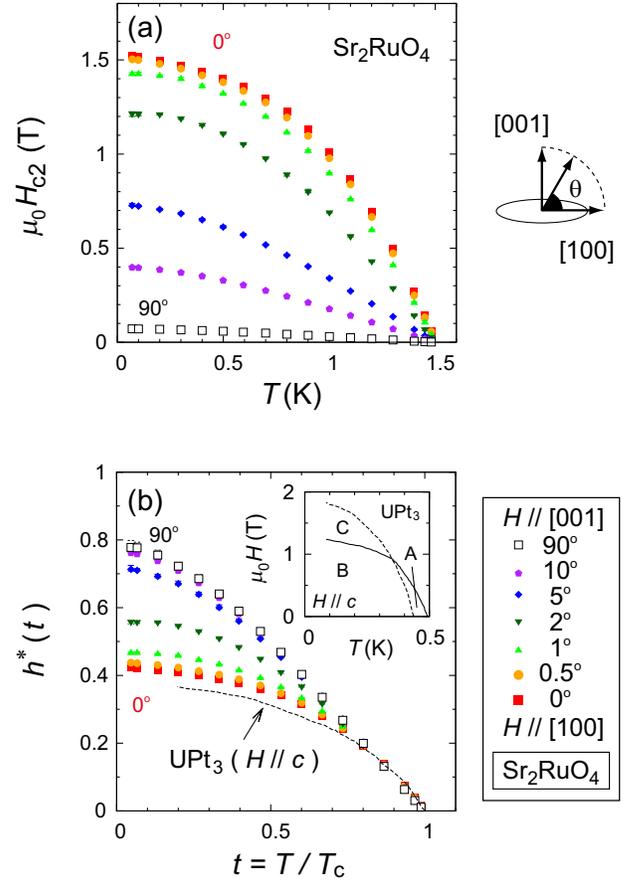}
\caption{
(Color online) (a) Field-temperature ($H$\,-\,$T$) phase diagram of $\214$ at 
$\theta=0^\circ, 0.5^\circ, 1^\circ, 2^\circ, 5^\circ, 10^\circ$, and $90^\circ$ from top to bottom for $\phi=0^\circ$.
(b) Temperature dependence of $h^*$ defined as eq.~\eqref{hstar}.
The inset represents the $H$\,-\,$T$ phase diagram of UPt$_3$ for $H \parallel c$ (Ref.~\onlinecite{Dijk1993JLTP}).
The dashed curve in (b) is $h^*(t)$ of the boundary of the ( B + C ) phase in UPt$_3$ for $H \parallel c$ (the dashed curve in the inset).
}
\label{HT}
\end{figure}

Figure \ref{HT}(a) is the field-temperature ($H$\,-\,$T$) phase diagram in various field directions at $\phi=0^\circ$,
where $\phi$ denotes the azimuthal angle within the $ab$ plane between the magnetic field and the [100] axis.
At $\phi=0^\circ$, no anomaly due to the mosaic structure was seen in the raw $\chi_\mathrm{ac}$ data, as shown in Fig.~\ref{raw}(a).
Reflecting the large anisotropy of $\Hc2$ in $\214$, $\Hc2$ becomes rapidly small
when the angle $\theta$ between the magnetic field and the $ab$ plane increases from 0$^\circ$.
In the specific heat measurements, the second superconducting transition was observed just below $\Hc2$ at low temperatures below 0.8 K. \cite{Deguchi2002JPSJ} 
Although such an additional transition was observed below 0.6 K in the ac susceptibility measurements, 
it was difficult to unambiguously identify it to be attributable to the second superconducting transition.\cite{Yaguchi2002PRB}
This is also the case for the present study.
One possible reason for this difficulty is that ac susceptibility, mainly probing the vortex movements, may not be sensitive 
to the small change in the entropy detected by specific heat measurements.
Therefore, we do not focus on the feature of the additional transition in this paper.

To characterize the limit of $\Hc2(T)$, 
we normalized $\Hc2$ by the initial slope at $\Tc$:
\begin{equation}
h^*(t) = - \frac{H_\mathrm{c2}(t)}{\mathrm{d}H_\mathrm{c2} / \mathrm{d}t|_{t=1}}\ \ \ \ \ \ (t \equiv T/T_\mathrm{c}). \label{hstar}
\end{equation}
If $\Hc2$ is determined by the orbital pair-breaking effect,
which originates from the kinetic energy of supercurrent around magnetic vortices,
$\Hc2$ is described by the Werthamer-Helfand-Hohenberg (WHH) theory \cite{Helfand1966PR,Werthamer1966PR} and its extension to $p$-wave superconductors.\cite{Maki1999JS,Lebed2000PhysicaC}
In these theories, it is expected that $h^*(t)$ increases linearly on cooling and is weakly suppressed at low temperatures with 
$h^*(t=0) \sim 0.7$.
In Fig.~\ref{HT}(b), we plot $h^*(t)$ with different $\theta$ at $\phi=0^\circ$.
The initial slope $\mathrm{d}H_\mathrm{c2} / \mathrm{d}t|_{t=1}$ is defined from the linear fit to $\Hc2(t)$ in the region $0.85 \le t \le 1$.
For $|\theta| \ge 5^\circ$, $h^*(t)$ behaves as expected from the WHH theory.
In contrast, for $|\theta| \le 2^\circ$, $h^*(t)$ is strongly limited at low temperatures. 
This result indicates that the $\Hc2$ limit in $\214$ is prominent for $|\theta| < 5^\circ$.
To emphasize the $\Hc2$ limit in another spin-triplet superconductor UPt$_3$, 
we plot, in Fig.~\ref{HT}(b) with the dashed curve, $h^*(t)$ of the boundary of the (~B~+~C~) phase for $H \parallel c$.
Although we chose the less limited one between the two $\Hc2(T)$ curves in UPt$_3$ for $H \parallel c$, 
a strong limit of $h^*(t)$ is clearly seen.

\begin{figure}
\includegraphics[width=2.5in]{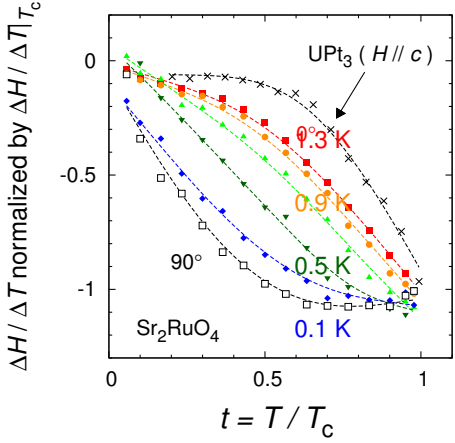}
\caption{
(Color online) Temperature dependence of the slope of the $H$-$T$ phase diagram of $\214$ at 
$\theta=90^\circ, 5^\circ, 2^\circ, 1^\circ, 0.5^\circ$, and $0^\circ$ from bottom to top for $\phi=0^\circ$.
The slope of the $\Hc2$ curve for UPt$_3$ (Ref.~\onlinecite{Dijk1993JLTP}), 
evaluated from the dashed curve in the inset of Fig.~\ref{HT}(b), is also plotted with crosses.
The dashed curves are guides to the eye.
}
\label{deri}
\end{figure}

If the orbital pair-breaking effect is mainly responsible for determining $\Hc2$, 
the slope of the $H$\,-\,$T$ phase diagram should be constant down to well below $\Tc$.
To identify the limit of $\Hc2$ near $\Tc$, 
we evaluate the slope at temperature $(T_1+T_2)/2$ as $\Delta \Hc2 / \Delta T = [\Hc2(T_1) - \Hc2(T_2)] / (T_1-T_2)$, 
where $T_1$ and $T_2$ are temperatures of adjacent data points.
The results are shown in Fig.~\ref{deri}.
For $|\theta| \ge 5^\circ$, the slope is constant down to approximately 1~K and approaches zero at low temperatures, 
which is well explained by the orbital pair-breaking effect.
However, for $|\theta| < 5^\circ$, the slope near $\Tc$ is not temperature-independent any more. 
This result suggests that 
the $\Hc2$ limit observed for $|\theta| < 5^\circ$ is present not only at low temperatures, but also at temperatures close to $\Tc$.
The slope of the $\Hc2(T)$ curve of UPt$_3$ for $H \parallel c$ (the dashed curve in Fig.~\ref{HT}(b)) also continues to vary up to $\Tc$, 
as plotted in Fig.~\ref{deri}. 

\begin{figure}
\includegraphics[width=3.2in]{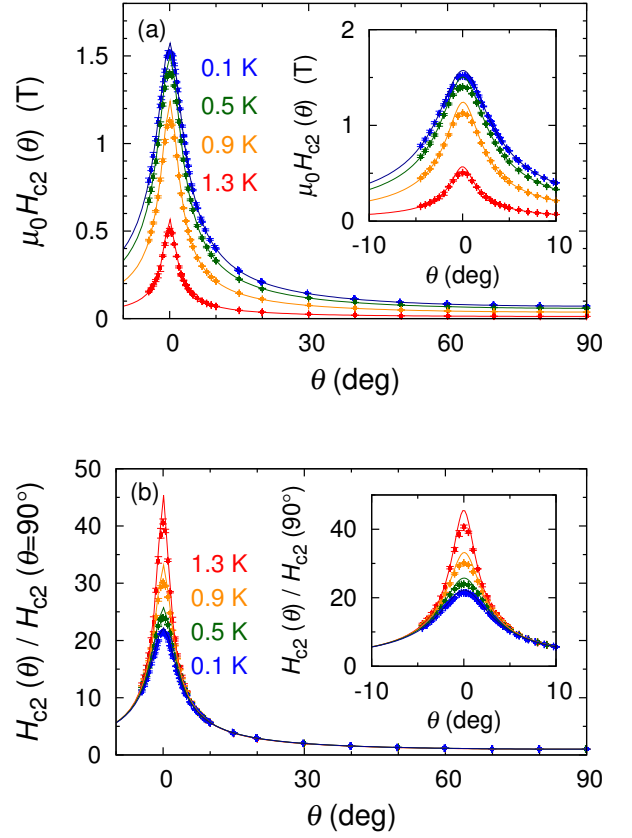}
\caption{
(Color online) (a) Field-angle $\theta$ dependence of the upper critical field $\Hc2$ of $\214$ at various temperatures for $\phi=0^\circ$ .
(b) Angle $\theta$ dependence of $\Hc2$ normalized by $\Hc2(\theta=90^\circ)$.
Solid curves are the fitting results using eq.~\eqref{Hc2p} in the range $2.5^\circ \le |\theta| \le 90^\circ$. 
Insets are enlarged views near $\theta=0^\circ$.
}
\label{theta}
\end{figure}

In order to characterize the non-linear temperature dependence of $\Hc2$ in $\214$,
we fitted $\Hc2(t)$ for $\theta=0^\circ$ by $a(1-t)^n$ with fitting parameters $a$ and $n$.
In any fitting range, $n$ is obviously larger than $n=0.5$,
which is expected for the two-dimensional (2D) superconductivity;\cite{Abrikosov}
the fitting in the range $0.9<t<1$ yields $n=0.9$.
In addition, the coherence length along the $c$ axis $\xi_c$ is estimated to be  3.2~nm using the GL equation
\begin{equation}
\xi_c=(\textit{\PHi}_0H_{\mathrm{c2} \parallel c}/2\pi H_{\mathrm{c2} \parallel ab}^2)^{1/2} \label{xi}
\end{equation}
with $\mu_0H_{\mathrm{c2} \parallel ab}=1.5$~T and $\mu_0H_{\mathrm{c2} \parallel c}=0.075$~T.
Here, $\textit{\PHi}_0$ is the flux quantum.
This value of $\xi_c$ is five times larger than the spacing of the conductive RuO$_2$ layers (0.62~nm).\cite{Mackenzie2003RMP}
Even if the WHH value $-0.7\mathrm{d}\Hc2/\mathrm{d}t|_{t=1}=2.5$~T is used for $H_{\mathrm{c2} \parallel ab}$,
$\xi_c=1.9$~nm is obtained.
These facts indicate that the superconductivity of $\214$ cannot be classified as a 2D superconductivity.

Figure \ref{theta}(a) represents the $\theta$ dependence of $\Hc2$ at various temperatures for $\phi=0^\circ$.
The $\theta$ dependence of $\Hc2$ normalized by $\Hc2(\theta=90^\circ)$ is also plotted in Fig.~\ref{theta}(b).
We found that, although $\Hc2(\theta)/\Hc2(90^\circ)$ for $10^\circ \le |\theta| \le 90^\circ$ is nearly independent of temperature, 
it decreases on cooling for $|\theta| < 10^\circ$.

\begin{figure}
\includegraphics[width=3.2in]{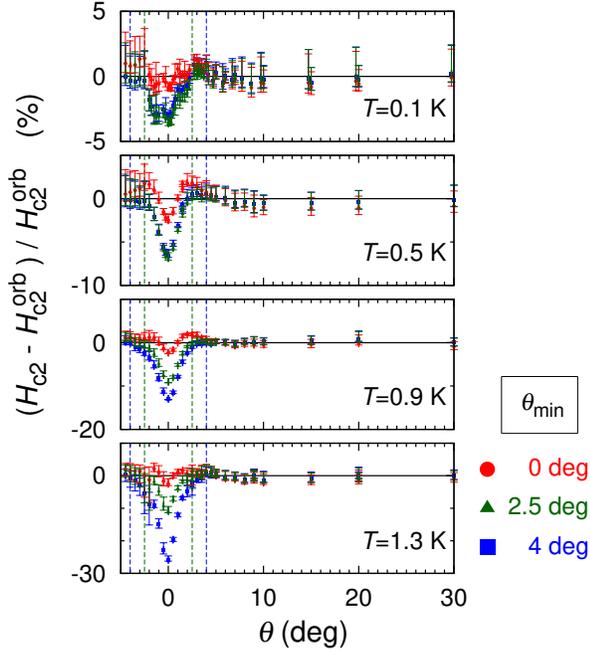}
\caption{
(Color online) Error of the fitting of eq.~\eqref{Hc2p} to the $\Hc2(\theta)$ data for $\phi=0^\circ$ 
in the range $\theta_\mathrm{min} \le |\theta| \le 90^\circ$ for several $\theta_\mathrm{min}$.
$\theta_\mathrm{min}=2.5^\circ$ is likely to give the most appropriate fitting range.
}
\label{fit}
\end{figure}

\begin{table}[b]
\begin{center}
\caption{
The upper critical field and its anisotropy. 
Fitting parameters are obtained by the fit of $\Hc2(\theta)$ for $\phi=0^\circ$  using eq.~\eqref{Hc2p} with $\theta_\mathrm{min}=2.5^\circ$ at each temperature.
}
\renewcommand{\arraystretch}{1.7}
\begin{tabular*}{8cm}{@{\extracolsep{\fill}}ccc|cc}\hline
\multicolumn{3}{c|}{Experiment} & \multicolumn{2}{c}{Fitting parameters} \\ \hline
$T$(K) & $\mu_0H_{\mathrm{c2} \parallel ab}$ & $H_{\mathrm{c2} \parallel ab}/H_{\mathrm{c2} \parallel c}$ & $\mu_0H_{\mathrm{c2} \parallel ab}^\mathrm{orb}$ & $\textit{\GAmma}$ \\ \hline
0.1 & 1.517~T & 21.4 & 1.574~T & 22.1 \\
0.5 & 1.399~T & 23.7 & 1.504~T & 25.5 \\
0.9 & 1.130~T & 30.5 & 1.243~T & 33.2 \\
1.3 & 0.496~T & 41.3 & 0.568~T & 46.1 \\
\hline
\end{tabular*}
\label{para}
\end{center}
\end{table}

We found that the observed $\Hc2(\theta)$ is well explained by the Ginzburg-Landau (GL) theory for 
anisotropic three-dimensional (3D) superconductors,\cite{Morris1972PRB}
if we allow the anisotropic ratio $\textit{\GAmma}=\Hc2(0^\circ)/\Hc2(90^\circ)$ to depend on temperature. 
The angular dependence of $\Hc2$ is expressed as
\begin{equation}
\Hc2^\mathrm{orb}(T,\theta) = \frac{H_{\mathrm{c2} \parallel ab}^\mathrm{orb}(T)}{\sqrt{\textit{\GAmma}(T)^2 \sin^2 \theta + \cos^2 \theta}}. \label{Hc2p}
\end{equation}
We fit eq.~\eqref{Hc2p} to the observed $\Hc2(\theta)$ at temperature $T_0$
with two fitting parameters $H_{\mathrm{c2} \parallel ab}^\mathrm{orb}(T_0)$ and $\textit{\GAmma}(T_0)$.
We chose the fitting range as $\theta_\mathrm{min} \le |\theta| \le 90^\circ$
so that the range is as wide as possible while the fitting yields a good result in the whole chosen range.
Figure~\ref{fit} represents the error of the fitting for different $\theta_\mathrm{min}$ at each temperature.
When $\theta_\mathrm{min} \ge 2.5^\circ$, $\Hc2(\theta)$ is well fitted by eq.~\eqref{Hc2p} in the chosen fitting range.
By contrast, for $\theta_\mathrm{min}<2.5^\circ$, $\Hc2(\theta)$ exhibits systematic deviation from eq.~\eqref{Hc2p} around $|\theta| \sim 2^\circ$.
Thus, we conclude that $\theta_\mathrm{min}=2.5^\circ$ is the most appropriate.
The fitting results are plotted in Fig.~\ref{theta} with the solid curves and the obtained fitting parameters are listed in Table~\ref{para}. 
Interestingly, the observed $\Hc2(\theta)$ is fitted by eq.~\eqref{Hc2p} better at lower temperatures,
as being clear in the inset of Fig.~\ref{theta}(b).
This tendency is also clear when the fit ratio $\textit{\GAmma}$ is compared with the experimental ratio $\Hc2(0^\circ)/\Hc2(90^\circ)$.
We should mention that the thin-film model \cite{Thinkham2nd4-10} applied to a 2D superconductor, \cite{Zuo2000PRB}
in which $\Hc2(\theta)$ exhibits a cusp at $\theta=0^\circ$, cannot account for our data.
While we carefully examined the $\theta$ dependence of $\Hc2$,
a kink in $\Hc2(\theta)$ around $\theta=2^\circ$ revealed by the specific heat measurements at 0.1~K (Ref.~\onlinecite{Deguchi2002JPSJ}) 
was not detected in the present study. 
We note that a kink in $\Hc2(\theta)$ was not detected in the thermal conductivity measurements at 0.32~K, either (Fig. 4(a) in Ref.~\onlinecite{Deguchi2002JPSJ}).
On the basis of the presently available results, we cannot clarify why the kink in $\Hc2(\theta)$ was observed only in the specific heat measurement at 0.1~K.

\begin{figure}
\includegraphics[width=3in]{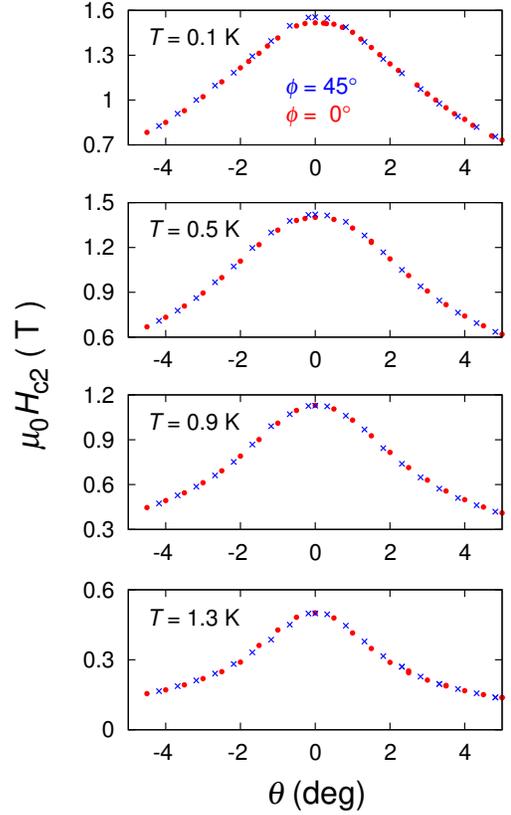}
\caption{
(Color online) Comparison between $\Hc2(\theta)$ at $\phi=0^\circ$ (circles) and $\phi=45^\circ$ (crosses).
}
\label{com}
\end{figure}

\begin{figure}
\includegraphics[width=2.6in]{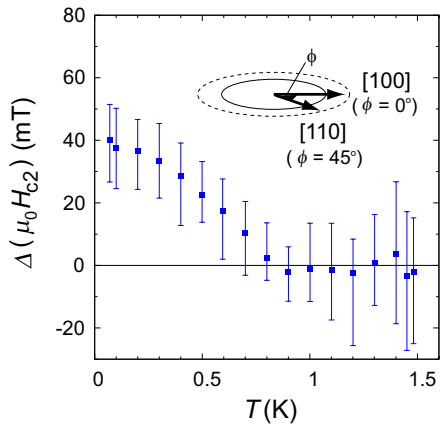}
\caption{
(Color online) Temperature dependence of the in-plane $\Hc2$ anisotropy between $\phi=0^\circ$ and $\phi=45^\circ$ at $\theta=0^\circ$.
$\Delta(\mu_0\Hc2)$ is defined as $\mu_0\Hc2(\phi=45^\circ)-\mu_0\Hc2(\phi=0^\circ)$.
}
\label{dif}
\end{figure}

In Fig.~\ref{com}, we compare the $\theta$ dependence of $\Hc2$ at angles between $\phi=0^\circ$ and $\phi=45^\circ$. 
As indicated by Fig.~\ref{raw}(b), two onset features appear in the field dependence of $\Delta\chi^\prime$ at $\phi=45^\circ$, 
reflecting the fact that the present sample consists mainly of two domains.
Since it is possible to separate the contribution from each of these domains, 
we plot in Fig~\ref{com} $\Hc2(\theta)$ for the major domain.
From Fig.~\ref{com}, we found that $\Hc2(T,\theta)$ at $\phi=45^\circ$ is both qualitatively and quantitatively similar to $\Hc2(T,\theta)$ at $\phi=0^\circ$.
Small difference in $\Hc2(\theta)$ at angles between $\phi=0^\circ$ and $\phi=45^\circ$ is observed only when the magnetic field is applied nearly parallel to the $ab$ plane.
Figure \ref{dif} represents the temperature dependence of the in-plane $\Hc2$ anisotropy between $\phi=0^\circ$ and $\phi=45^\circ$.
Here, we define $\Delta(\mu_0\Hc2)$ as $\mu_0\Hc2(\phi=45^\circ)-\mu_0\Hc2(\phi=0^\circ)$.
Although the temperature $T^*$ at which $\Delta (\mu_0\Hc2)$ starts to increase on cooling depends on samples (0.9~K $\le T^* \le 1.2$~K),
$\Delta(\mu_0 \Hc2)$ of 40~mT at low temperatures is nearly the same among different samples with best $\Tc$.\cite{Mao2000PRL}

On the basis of the phenomenological theory proposed by Gorkov,\cite{Gorkov1987SSR} 
superconductivity with a two-component order parameter, $k_x \pm ik_y$, 
should be accompanied by a substantial four-fold anisotropy in the in-plane $\Hc2$.
However, as presented in Fig.~\ref{dif}, no in-plane $\Hc2$ anisotropy is observable above about 1~K; 
the anisotropy grows on cooling, but reaches at most 3\% at low temperatures.
This lack of the large in-plane $\Hc2$ anisotropy is attributable to the multiband effect.\cite{Agterberg2001PRB,Kusunose2004JPSJ,Mineev2008PRB}
Because the directions of the gap minima are 45 degrees different between the active ($\gamma$) and passive ($\alpha$ and $\beta$) bands, \cite{Nomura2005JPSJ}
the $\Hc2$ anisotropy reflecting the gap structure on different Fermi surface sheets can be cancelled. \cite{Agterberg2001PRB,Kusunose2004JPSJ}

\section{Discussion}

Let us discuss the origin of the $\Hc2$ limit in $\214$.
For both 2D and 3D superconductors in which the main pair-breaking effect is due to the ordinary orbital effect,
such a limit is not expected.
Thus, in order to explain the $\Hc2$ limit, we need an additional pair-breaking mechanism.

One of the possible additional pair-breaking effects in $\214$ is an unusual orbital pair-breaking effect.
For example, in a nearly 2D superconductor (TMET-STF)$_2$BF$_4$,\cite{Uji2001PRB} 
it is proposed that $\Hc2(T)$ for $H \parallel a$ is limited due to the limit of the coherence length by the layer spacing,
which leads to the decrease of the anisotropy ratio of $\Hc2$ on cooling.
For $\214$, $\xi_c$ estimated using eq.~\eqref{xi}, for which the ordinary orbital pair-breaking effect is assumed,
is limited to be approximately 3.2~nm below 1~K. 
In fact, if we strictly apply eq.~\eqref{xi},
$\xi_c$ takes a minimum at 0.8~K and even increases by about 3\% at low temperatures.
However, we cannot find a clear answer to this limit because 
the limited value of $\xi_c$, 3.2 nm, is five times larger than the layer spacing. 
Therefore, the origin of the apparent limit of the coherence length in $\214$ seems different from that in (TMET-STF)$_2$BF$_4$.

Recently, Machida and Ichioka proposed the Pauli effect 
as an additional pair-breaking effect leading to the $\Hc2$ limit in $\214$.\cite{Machida2008PRB}
Using a model with a single-band spherical Fermi surface and by assuming the Pauli effect,
they reproduced the observed $H_\mathrm{c2}(\theta)$ at 0.1 K (Ref.~\onlinecite{Deguchi2002JPSJ}) 
as well as field dependences of the specific heat \cite{Deguchi2004JPSJ} and magnetization.\cite{Tenya2006JPSJ}
Interestingly, we found that the Machida-Ichioka model well reproduces our results of $\Hc2(T,\theta)$, too. 
Nevertheless, this would not lead to the conclusion that the $\Hc2$ limit in $\214$ is attributable to the Pauli effect
because the Machida-Ichioka model overlooks some key experimental as well as theoretical facts.
First, Machida-Ichioka model does not include the multiband effect. 
Their single-band model explains the field dependence of the specific heat at low temperatures.
However, $\214$ has three cylindrical Fermi surfaces, $\alpha$, $\beta$, and $\gamma$.\cite{Mackenzie2003RMP,Bergemann2003AP}
Although the active band $\gamma$ is dominant in the superconductivity in high fields,
the passive bands $\alpha$ and $\beta$ also contribute to the superconductivity in low fields.\cite{Deguchi2004PRL}
The contribution from the $\alpha$ and $\beta$ bands is essential to explain the plateau-like dependence quantitatively.\cite{Deguchi2004PRL}
In fact, inclusion of the multiband effect is needed to explain the $T^2$ dependence of the specific heat at low temperatures in zero field.\cite{Agterberg1997PRL,Zhitomirsky2001PRL,Nomura2002JPSJ}
Secondly, as mentioned in Sec. I,
the Pauli effect contradicts the results of the Knight shift experiments.\cite{Ishida1998Nature,Ishida2001PRB,Duffy2000PRL}
Although they proposed the possibility that 
the spin part of the Knight shift was too small to be detected in the NMR experiments, 
the spin part at the Ru site is in reality as large as 4\%.\cite{Ishida2001PRB}
In addition, the superconductivity was distinctly observed in $1/T_1$ through Ru NMR in the identical setup.\cite{Murakawa2007JPSJ}
These facts exclude the possibility of the Pauli mechanism.
Therefore, an alternative mechanism needs to be introduced to explain both the Knight shift behavior and the $\Hc2$ limit.

\section{summary}

We have clarified the temperature and field-angle dependence of $\Hc2$ of $\214$.
Our experiments were performed with an accurate and precise control of the applied magnetic field to avoid errors due to the misalignment.
We revealed that the $\Hc2$ limit is clearly observed for $|\theta| < 5^\circ$ and
it occurs not only at low temperatures but also at temperatures close to $\Tc$.
We also found that, by assuming a temperature-dependent anisotropic ratio,
the GL theory for an anisotropic 3D superconductor can explain the angular dependence of $\Hc2$ well, particularly at lower temperatures.
The observed behavior of $\Hc2(\theta)$ is qualitatively the same between $\phi=0^\circ$ and $\phi=45^\circ$.
Only a small in-plane $\Hc2$ anisotropy was observed at low temperatures, 
which disappears rapidly as the magnetic-field direction leaves from the $ab$ plane.
Until now, the origin of the effective pair-breaking effect, 
which is compatible with both the invariance to the Knight shift and the limiting behavior of $\Hc2$, remains unclear.

\acknowledgments
We thank K. Machida, M. Ichioka, R. Ikeda, H. Ikeda, K. Deguchi, H. Yaguchi, Y. Nakai, H. Takatsu and M. Kriener for useful discussions and supports. 
This work is supported by a Grant-in-Aid for Global COE program ``The Next Generation of Physics, Spun from Universality and Emergence'' 
from the Ministry of Education, Culture, Sports, Science, and Technology (MEXT) of Japan.
It is also supported by Grants-in-Aid for Scientific Research from MEXT and from the Japan Society for the Promotion of Science (JSPS).
One of the authors (S. K.) is financially supported by JSPS.

\end{document}